\documentclass[11pt]{article}      

\usepackage{epsfig}

\newcommand{\ba}{\begin{eqnarray}}
\newcommand{\ea}{\end{eqnarray}}
\begin{document}

\title{Shape-phase transitions in nuclei and random interactions
\footnote{Lecture notes: IV International Balkan School on Nuclear 
Physics, Bodrum, Turkey, September 22-29, 2004}}

\author{Roelof Bijker\\
ICN-UNAM, AP 70-543, 04510 M\'exico D.F., M\'exico\\
E-mail: bijker@nuclecu.unam.mx}
\date{}
\maketitle

\begin{abstract}
In these lecture notes I present a short review of nuclear shapes, shape 
coexistence and shape-phase transitions in the interacting boson model. 
In a study with random interactions it is shown that the appearance of 
regular spectral features is a far more common phenomenon than was previously 
thought. The origin of these features are explained by studying the relation 
with the underlying geometric shapes.
\end{abstract}

\section{Introduction}

The concept of shape coexistence and shape-phase transitions in atomic nuclei 
is currently a topic of active research both experimentally and theoretically 
\cite{nature,QPT}. 
The phenomenon of shape coexistence refers to the occurrence of 
different shapes in a given nucleus and has been observed in fission isomers 
in the actinides, intruder orbitals in or near closed (sub)shell nuclei and 
superdeformed nuclei at high spins \cite{coex}. 
Whereas coexisting shapes are usually 
associated with different configurations ({\it e.g.} normal and intruder), 
shape-phase transitions describe how one geometric shape evolves into 
another within the same configuration, {\it e.g.} between spherical and 
deformed shapes in the Nd, Sm and Gd nuclei or between $\gamma$-soft and 
deformed in the Pt-Os mass region \cite{DSI}.

It is the aim of these lecture notes to present a short overview of 
shape-phase transitions in the interacting boson model (IBM) and its 
extensions and to discuss the appearance of regular spectral features from 
the IBM with random interactions and its relation with the underlying 
geometric shapes and critical points. In the first lecture (Section 2), 
I present a review of shape-phase transitions for the IBM including two 
more recent developments, {\it i.e.} two-neutron transfer reactions as a probe 
of shape-phase coexistence in the Ge isotopes and the shape-phase diagram 
for the neutron-proton IBM. In the second lecture, I discuss some surprising 
results that were obtained in studies of the IBM with random interactions: 
the predominance of ground states with angular momentum $L = 0$ and the 
occurrence of both vibrational and rotational structures despite the random 
nature of the interactions.

\section{Shape-phase transitions in nuclei}

Shape-phase transitions in atomic nuclei were studied extensively in the 
early 80's in the framework of the IBM. The general procedure was laid out 
by Gilmore in the 70's using atomic coherent states \cite{cs} in combination 
with catastrophe theory \cite{Gilmore} which was applied to nuclear physics 
by Dieperink, Scholten and Iachello \cite{DSI}, Feng, Gilmore and Deans 
\cite{Feng} and L\'opez-Moreno and Casta\~nos \cite{ELM}. 
After the introduction of critical point symmetries as special solutions 
to the Bohr Hamiltonian \cite{FI,RFC}, the subject of shape-phase transitions 
has witnessed a revival both experimentally and theoretically. For a recent 
review of shape-phase transitions and critical point symmetries 
in nuclear physics, see \cite{QPT}. 

\subsection{The interacting boson model}

The IBM provides an elegant and 
powerful tool for the description of collective nuclei \cite{IBM}.  
In this model, collective excitations in nuclei are described in terms of a 
system of $N$ interacting monopole ($s^{\dagger}$) and quadrupole 
($d^{\dagger}_m$ with $m=0,\pm1,\pm2$) bosons with $L^P=0^+$ and $2^+$, 
respectively. The the number of bosons $N$ is determined by half the number 
of valence nucleons. In addition to $N$, the 
eigenfunctions have good angular momentum and parity $L^P$. 
Let's consider the schematic IBM Hamiltonian of 
the consistent-Q formulation (CQF) \cite{CQF} 
\ba
H \;=\; \epsilon \, \hat n_d - \kappa \, \hat Q(\chi) \cdot \hat Q(\chi) ~, 
\label{hcqf}
\ea
where $\hat n_d$ is the number operator for quadrupole bosons
\ba
\hat n_d \;=\; \sqrt{5} \, (d^{\dagger} \times \tilde{d})^{(0)} ~,
\ea
and $\hat Q(\chi)$ denotes the quadrupole operator
\ba
\hat Q_{m}(\chi) \;=\; (s^{\dagger} \times \tilde{d} 
+ d^{\dagger} \times \tilde{s})^{(2)}_m  
+ \chi \, (d^{\dagger} \times \tilde{d})^{(2)}_m ~.
\ea
Here $\tilde{s}=s$ and $\tilde{d}_m=(-1)^{2-m}d_{-m}$. 
The Hamiltonian of Eq.~(\ref{hcqf}) describes the main features of 
collective nuclei: it contains the dynamical symmetries of the IBM for 
special choices of the coefficients $\epsilon$, $\kappa$ and $\chi$, and 
allows to describe the transitional regions between any of symmetry limits 
as well. 

\subsection{Potential-energy surface}
\label{pes}

The connection between the IBM, potential-energy surfaces, geometric 
shapes and phase transitions can be investigated by introducing a 
coherent, or intrinsic, state which is expressed as a boson condensate 
\cite{DSI}
\ba
\left| N,\beta,\gamma \right> \;=\; \frac{1}{\sqrt{N!}} 
\left( b_c^{\dagger} \right)^n \left| 0 \right> ~,
\label{cost}
\ea
with
\ba
b_c^{\dagger} \;=\; \frac{1}{\sqrt{1+\beta^2}} \left( s^{\dagger} 
+ \beta \cos \gamma \, d_0^{\dagger} + \frac{1}{\sqrt{2}} \beta 
\sin \gamma ( d_2^{\dagger} + d_{-2}^{\dagger} ) \right) ~,
\label{condensate}
\ea
where the variables $\beta$ and $\gamma$ determine the geometry of 
the nuclear surface. Spherical shapes are characterized by $\beta = 0$ and 
deformed ones by $\beta > 0$. The angle $\gamma$ allows one to distinguish 
between axially deformed nuclei, $\gamma = 0^{\circ}$ for prolate and 
$\gamma = 60^{\circ}$ for oblate deformation, and triaxial nuclei 
$0^{\circ} < \gamma < 60^{\circ}$.

The potential-energy surface is given by the expectation value 
of the Hamiltonian in the coherent state \cite{DSI}
\ba
V(\beta,\gamma) &=& 
\left< N,\beta,\gamma \right| :H: \left| N,\beta,\gamma \right> 
\nonumber\\
&=& \frac{\epsilon N \beta^2}{1+\beta^2} 
- \frac{\kappa N(N-1)}{(1+\beta^2)^2} \left[ 4 \beta^2 
- 4 \chi \sqrt{\frac{2}{7}} \beta^3 \cos 3\gamma 
+ \frac{2}{7} \chi^2 \beta^4 \right] ~. 
\label{eibm1}
\ea
The normal ordered form of the Hamiltonian $:H:$ was used to eliminate the 
contributions of the one-body terms of the quadrupole-quadrupole interaction 
to the energy surface $V(\beta,\gamma)$. 

The equilibrium shape associated to the IBM Hamiltonian 
can be obtained by determining the minimum of the energy surface with 
respect to the geometric variables $\beta$ and $\gamma$, {\it i.e.} the first 
derivatives vanish
\ba
\frac{\partial V}{\partial \beta} \;=\; 
\frac{\partial V}{\partial \gamma} \;=\; 0 ~, 
\label{der1}
\ea
and the determinant of the Hessian matrix of the second derivatives 
is positive
\ba 
\left( \frac{\partial^2 V}{\partial \beta^2} \right)
\left( \frac{\partial^2 V}{\partial \gamma^2} \right)
- \left( \frac{\partial^2 V}{\partial \beta \partial \gamma} \right)^2 > 0 ~.
\label{der2}
\ea

\subsection{Dynamical symmetries} 

First some general properties of the IBM are reviewed, in particular  
the equilibrium shapes for the dynamical symmetries that arise for 
special choices of the coefficients $\epsilon$, 
$\kappa$ and $\chi$ in the Hamiltonian of Eq.~(\ref{hcqf}). 
A more detailed discussion can be found in \cite{IBM}. 

\subsection*{The $U(5)$ limit}

For $\kappa=0$, the Hamiltonian $H$ of Eq.~(\ref{hcqf}) reduces to the 
$U(5)$ limit of the IBM 
\ba
H_1 \;=\; \epsilon \, \hat n_d ~. 
\label{h1}
\ea
The potential-energy surface of $H_1$ is given by 
\ba
V_1(\beta) \;=\; \frac{\epsilon N \beta^2}{1+\beta^2} ~. 
\label{v1}
\ea
The equilibrium value of the deformation parameter $\beta$ is 
easily obtained by solving Eqs.~(\ref{der1}) and~(\ref{der2}) 
to give $\beta_e = 0$ which corresponds to a spherical shape. 

\subsection*{The $SO(6)$ limit}
 
For $\epsilon=0$ and $\chi=0$, one recovers the the $SO(6)$ limit 
\ba
H_2 \;=\; - \kappa \, \hat Q(\chi) \cdot \hat Q(\chi) ~, \hspace{1cm} 
\chi \;=\; 0 ~.
\label{h2}
\ea
Also in this case, the energy surface only depends on $\beta$ 
\ba
V_2(\beta) \;=\;
- \frac{4 \kappa N(N-1) \beta^2}{(1+\beta^2)^2} ~. 
\label{v2}
\ea
The equilibrium value is given by $\beta_e = 1$   
corresponding to a $\gamma$-unstable deformed shape. 

\subsection*{The $SU(3)$ limit}
 
For $\epsilon=0$ and $\chi=\mp \sqrt{7}/2$, the schematic Hamiltonian 
of Eq.~(\ref{hcqf}) reproduces the $SU(3)$ limit 
\ba
H_3 \;=\; - \kappa \, \hat Q(\chi) \cdot \hat Q(\chi) ~, \hspace{1cm} 
\chi=\mp \frac{1}{2} \sqrt{7} ~.
\label{h3}
\ea
For this case, the potential-energy surface depends both on $\beta$ and 
$\gamma$ 
\ba
V_3(\beta,\gamma) \;=\;
- \frac{\kappa N(N-1)}{(1+\beta^2)^2} \left[ 4 \beta^2 
\pm 2 \sqrt{2} \beta^3 \cos 3\gamma 
+ \frac{1}{2} \beta^4 \right] ~. 
\label{v3}
\ea
The equilibrium values are given by $\beta_e = \sqrt{2}$ and 
$\gamma_e = 0^{\circ}$ for $\chi=-\frac{1}{2}\sqrt{7}$ and by 
$\beta_e = \sqrt{2}$ and $\gamma_e = 60^{\circ}$ for 
$\chi=\frac{1}{2}\sqrt{7}$ 
corresponding to prolate and oblate deformed shapes, respectively. 

\subsection{Shape-phase transitions}
\label{shapes}

The transitional regions between any two of the dynamical symmetries of 
the IBM can be studied by combining the corresponding Hamiltonians, and 
varying the relative strength of the interaction terms. In 1980, 
Dieperink, Scholten and Iachello showed that these transitional regions 
exhibit shape-phase transitions \cite{DSI}. 

In 1996, the subject was revisited by L\'opez-Moreno and Casta\~nos 
\cite{ELM} who investigated the shape-phase transitions for general (one- 
and two-body) IBM Hamiltonians using the full power of catastrophy theory 
\cite{catastrophe}. 
It was shown that the shape-phase diagram depends on two independent 
combinations of the parameters of the IBM Hamiltonian, called $r_1$ and 
$r_2$, which can be used to classify the equilibrium configurations. 
For the Hamiltonian of Eq.~(\ref{hcqf}) they are given by
\ba
r_1 \;=\; \frac{-4+\frac{\epsilon}{\kappa(N-1)}}
{-\frac{4}{7} \chi^2+4+\frac{\epsilon}{\kappa(N-1)}} ~,
\hspace{1cm} 
r_2 \;=\; \frac{-8 \chi \sqrt{\frac{2}{7}}}
{-\frac{4}{7} \chi^2+4+\frac{\epsilon}{\kappa(N-1)}} ~.
\label{control}
\ea
In addition, closed analytic expressions were derived for the separatrix, 
which consists of the critical points
\ba
r_{1c} \;=\; -\frac{1}{2} + \frac{1}{2} \sqrt{1+\frac{1}{2}r_2^2} ~,
\label{r1c}
\ea
the spinodal and antispinodal points
\ba
r_{1s} \;=\; -1 + \frac{(9r_2^2+16)^{3/2}-64}{54 r_2^2} ~, 
\hspace{1cm} r_{1a} \;=\; 0 ~.
\label{r1a}
\ea
In Table~\ref{values} the values of $r_1$ and $r_2$ are given 
for each of the dynamical symmetries discussed in the previous section. 

\begin{table}
\centering
\caption[Equilibrium values]
{\small Equilibrium values of the deformation parameters 
for the dynamical symmetry limits of the IBM.}
\label{values}
\vspace{15pt}
\begin{tabular}{cccccccc}
\hline
& & & & & & \\
Limit & $\epsilon$ & $\kappa$ & $\chi$ 
& $r_1$ & $r_2$ & $\beta_e$ & $\gamma_e$ \\ 
& & & & & & \\
\hline
& & & & & & \\
$U(5)$ & $\epsilon$ & 0 & -- & 1 & 0 & 0 & -- \\
& & & & & & \\
$SO(6)$ & 0 & $\kappa$ & 0 & $-1$ & 0 & $\pm 1$ & -- \\
& & & & & & \\
$SU(3)$ & 0 & $\kappa$ & $-\frac{1}{2}\sqrt{7}$ 
& $-\frac{4}{3}$ & $ \frac{4}{3}\sqrt{2}$ & $\sqrt{2}$ & $ 0^{\circ}$ \\
& & & & & & \\
      & 0 & $\kappa$ & $+\frac{1}{2}\sqrt{7}$ 
& $-\frac{4}{3}$ & $-\frac{4}{3}\sqrt{2}$ & $\sqrt{2}$ & $60^{\circ}$ \\
& & & & & & \\
\hline
\end{tabular}
\end{table}

The methods used in \cite{DSI} and \cite{ELM} are closely related and 
give complementary results, which lead to a very simple and clear picture 
of the shape-phase diagram in nuclei as will be shown in the remainder 
of this section.  

\subsection*{The $U(5) - SO(6)$ transitional region}

The transition between the spherical and $\gamma$-unstable 
deformed shapes can be studied by considering the Hamiltonian
\ba
H_{12} \;=\; H_1 + H_2 \;=\; \epsilon \, \hat n_d 
- \kappa \, \hat Q(0) \cdot \hat Q(0) ~.
\label{h12}
\ea
The corresponding energy surface can be written as 
\ba
{\cal E}(\beta) \;=\; \frac{V_1(\beta)+V_2(\beta)}{N\epsilon} 
\;=\; \frac{\beta^2}{1+\beta^2} 
- 4\eta \frac{\beta^2}{(1+\beta^2)^2} ~, 
\label{v12}
\ea
where the energy functionals of Eqs.~(\ref{v1}) and~(\ref{v2}) were scaled 
by $\epsilon$ and the number of bosons $N$. The transition is described 
by varying the control parameter $\eta=\kappa(N-1)/\epsilon$ on the interval 
$0 \leq \eta < \infty$. The equilibrium value of $\beta$ depends on $\eta$ 
\ba
\eta \leq \frac{1}{4} &:& \hspace{1cm} \beta_e \;=\; 0 ~, 
\nonumber\\
\eta \geq \frac{1}{4} &:& \hspace{1cm} 
\beta_e^2 \;=\; \frac{4\eta-1}{4\eta+1} ~, 
\ea
{\it i.e.} a spherical minimum for $\eta \leq 1/4$ and a deformed one 
for $\eta > 1/4$. The phase transition takes place at the 
critical point $\eta_c = 1/4$. Its nature can 
be determined by studying the ground state energy 
${\cal E}_{\rm min}(\eta)={\cal E}(\beta_e)$ and its derivatives 
at the critical point \cite{Gilmore}: if the ground state energy is 
discontinuous in the control parameter $\eta$ 
the transitition is zeroth order, and if the first (second) 
derivative is discontinuous it is first (second) order.  
The ground state energy of Eq.~(\ref{v12}) is 
\ba
\eta \leq \frac{1}{4} &:& \hspace{1cm} {\cal E}_{\rm min}(\eta) \;=\; 0 ~,
\nonumber\\
\eta \geq \frac{1}{4} &:& \hspace{1cm} {\cal E}_{\rm min}(\eta) \;=\; 
-\frac{(4\eta-1)^2}{16\eta} ~.
\ea
The first derivative is continuous at $\eta_c=1/4$, whereas the second 
derivative is discontinous. Therefore, the $U(5)$-$SO(6)$ transitional 
region exhibits a second-order phase transition \cite{DSI}. 

\begin{figure}
\vfill 
\begin{minipage}{.5\linewidth}
\centerline{\epsfig{file=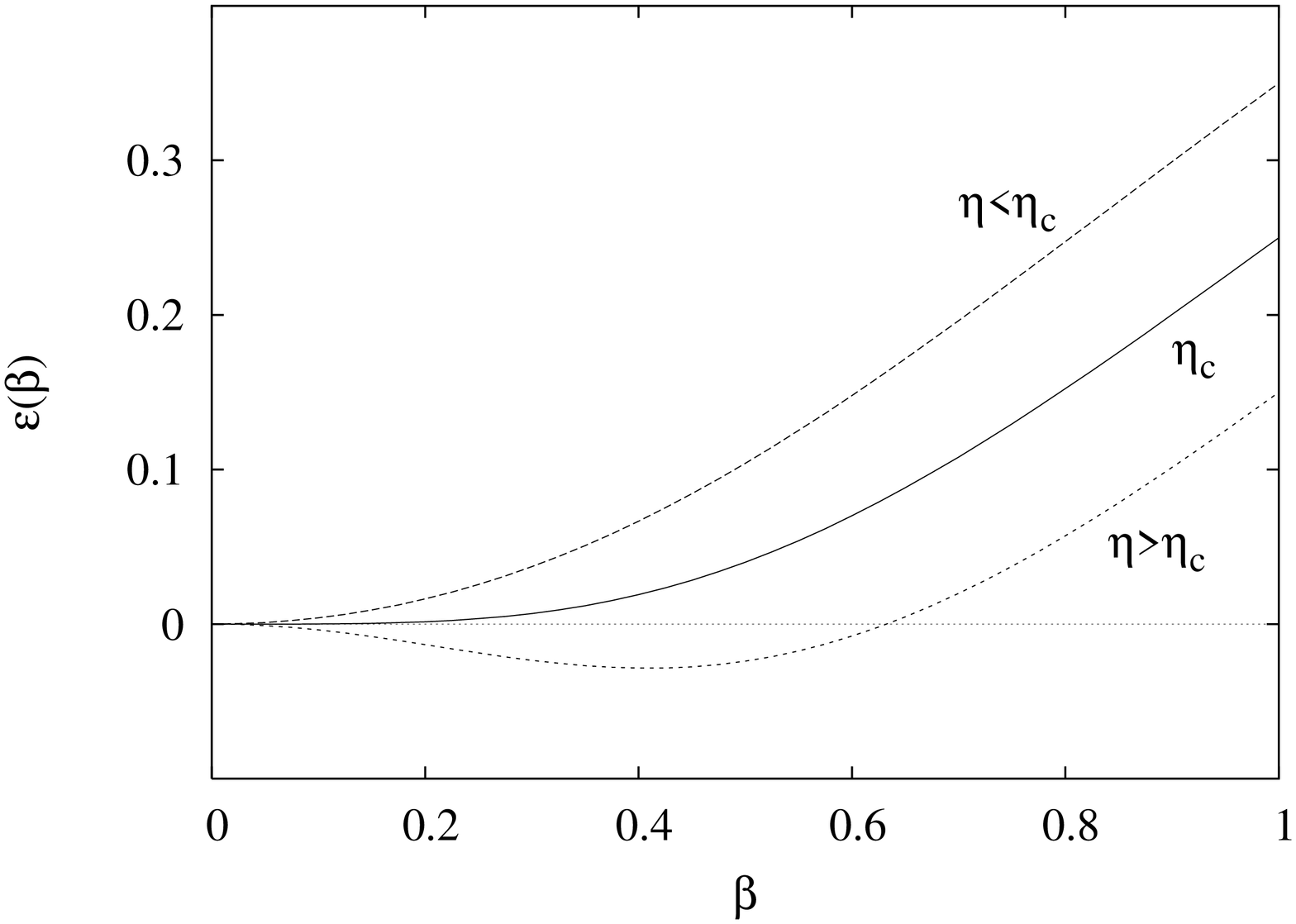,angle=270,width=\linewidth}}
\end{minipage}\hfill
\begin{minipage}{.5\linewidth}
\centerline{\epsfig{file=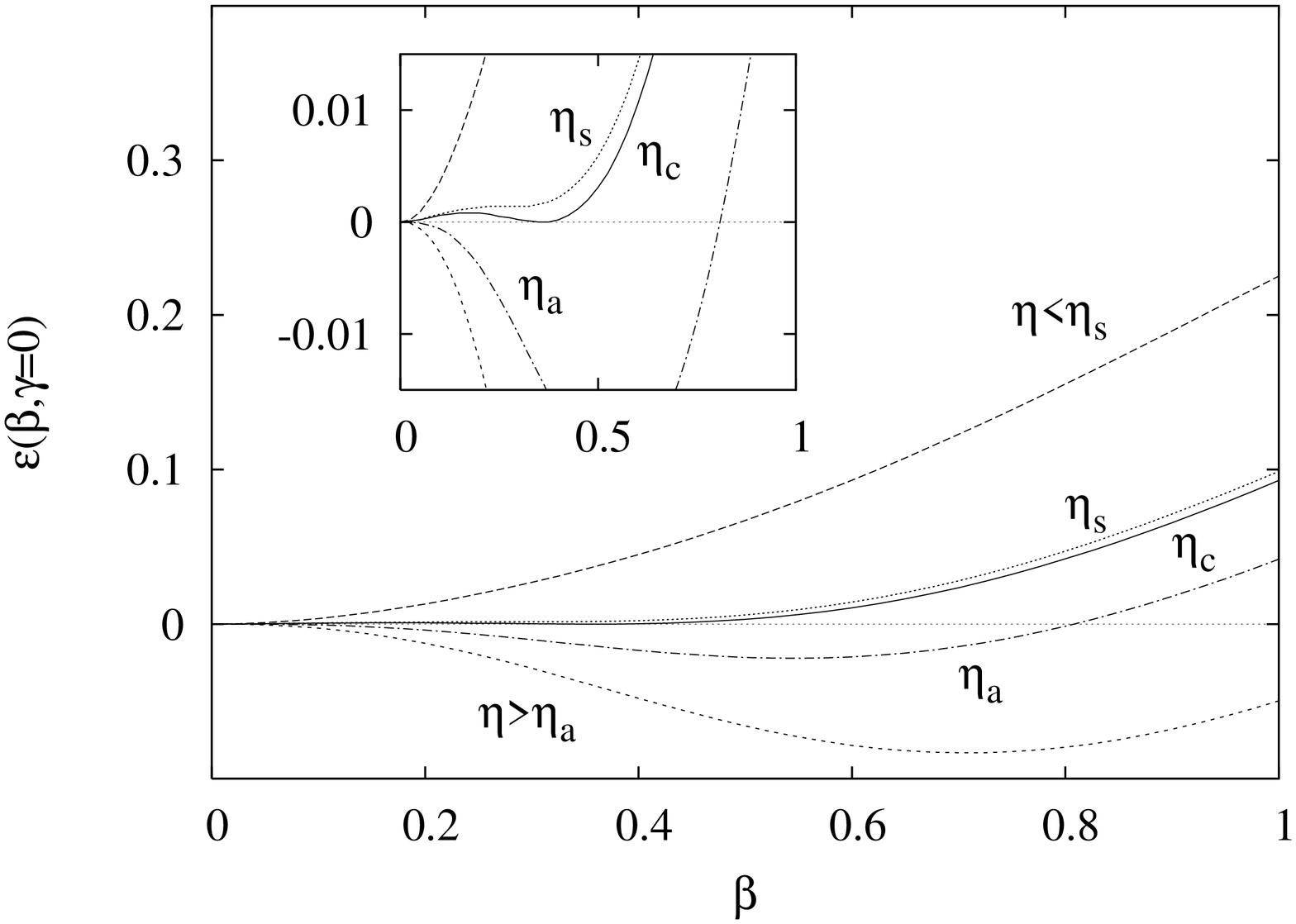,angle=270,width=\linewidth}}
\end{minipage}
\caption[Potential-energy surfaces]
{\small Potential-energy surfaces. (left) The second-order $U(5)-SO(6)$ 
phase transition: $\eta < \eta_c$ (dashed), $\eta = \eta_c$ (solid) and 
$\eta > \eta_c$ (short-dashed). (right) The first-order $U(5)-SU(3)$ 
phase transition: $\eta < \eta_s$ (dashed), $\eta = \eta_s$ (dotted), 
$\eta = \eta_c$ (solid), $\eta = \eta_a$ (dotted-dashed) and 
$\eta > \eta_a$ (short-dashed).}
\label{pot1}
\end{figure}

An analysis in terms of catastrophy theory gives that the values of 
$r_1$ and $r_2$ for the Hamiltonian of Eq.~(\ref{h12}) are given by
\ba
r_1 \;=\; \frac{1-4\eta}{1+4\eta} ~, \hspace{1cm}
r_2 \;=\; 0 ~.
\label{trans12}
\ea
In the $r_2$-$r_1$ plane the transitional region for $0 \le \eta < \infty$ 
is represented by $-1 \leq r_1 \leq 1$ and $r_2=0$. The critical, 
antispinodal and spinodal points are obtained by taking the intersection 
with the curves corresponding to Eqs.~(\ref{r1c}) and (\ref{r1a})  
\ba
\eta_c \;=\; \frac{1}{4} ~, &\hspace{1cm}& r_{1c} \;=\; r_{2c} \;=\; 0 ~, 
\nonumber\\
\eta_a \;=\; \frac{1}{4} ~, &\hspace{1cm}& r_{1a} \;=\; r_{2a} \;=\; 0 ~, 
\nonumber\\
\eta_s \;=\; \frac{1}{4} ~, &\hspace{1cm}& r_{1s} \;=\; r_{2s} \;=\; 0 ~. 
\ea
This shows that the critical, antispinodal and spinodal points coincide 
which is a signature for a second-order phase transition, in agreement with 
the results obtained above from the analysis of the ground state energy 
\cite{DSI}. In the left panel of Fig.~\ref{pot1}, the potential-energy 
surface is given for three different values of the control parameter $\eta$. 
For $\eta<\eta_c$ the equilibrium shape is spherical and for $\eta>\eta_c$ 
it is deformed. There is no region of coexistence of spherical and deformed 
minima. 

\subsection*{The $U(5) - SU(3)$ transitional region}

The transitional region between the spherical and axially deformed nuclei 
can be studied by considering the schematic Hamiltonian 
\ba
H_{13} \;=\; H_1 + H_3 \;=\; \epsilon \, \hat n_d 
- \kappa \, \hat Q(\chi) \cdot \hat Q(\chi) ~, \hspace{1cm} 
\chi=\mp \frac{1}{2} \sqrt{7} ~, 
\label{h13}
\ea
as a function of the control parameter $\eta=\kappa(N-1)/\epsilon$ with 
$0 \leq \eta < \infty$. The corresponding energy surface is given by
\ba
{\cal E}(\beta,\gamma) \;=\; \frac{V_1(\beta)+V_3(\beta,\gamma)}{N\epsilon} 
\;=\; \frac{\beta^2}{1+\beta^2} - \frac{\eta}{(1+\beta^2)^2} 
\left[ 4 \beta^2 \pm 2 \sqrt{2} \beta^3 \cos 3\gamma 
+ \frac{1}{2} \beta^4 \right] ~,
\label{v13}
\ea
where the energy functionals of Eqs.~(\ref{v1}) and~(\ref{v3}) are scaled 
by $\epsilon$ and the number of bosons $N$. The analysis of the ground 
state energy as a function of $\eta$ is more complicated than in the 
previous case due to the presence of the $\beta^3 \cos 3\gamma$ term.  
It was found that $H_{13}$ exhibits a first-order phase 
transition at the critical point $\eta_c=2/9$ \cite{DSI}. 

An analysis with catastrophe shows that the values of the coefficients 
$r_1$ and $r_2$ for $H_{13}$ are given by
\ba
r_1 \;=\; \frac{1-4\eta}{1+3\eta} ~, \hspace{1cm}
r_2 \;=\; \frac{\pm 4\sqrt{2}\eta}{1+3\eta} ~,
\ea
from which it is easy to see that the transitional region between the $U(5)$ 
and $SU(3)$ limits is characterized by a straight line 
\ba
r_1 \;=\; 1 \mp \frac{7}{4\sqrt{2}} r_2 ~, \hspace{1cm} 
-\frac{4}{3} \leq r_1 \leq 1 ~.
\label{trans13}
\ea
The critical, antispinodal and spinodal points are obtained 
by taking the intersection of the straight line of Eq.~(\ref{trans13}) 
with the curves corresponding to Eqs.~(\ref{r1c}) and (\ref{r1a}). 
The critical point is characterized by 
\ba
\eta_c \;=\; \frac{2}{9} ~, \hspace{1cm} 
r_{1c} \;=\; \frac{1}{15} ~, \hspace{1cm} 
r_{2c} \;=\; \pm \frac{8\sqrt{2}}{15} ~.
\ea
The equilibrium value of the deformation parameter at the critical point 
is given by \cite{ELM}
\ba
\beta_c \;=\; \frac{r_{2,c}}{1+\sqrt{1+r^2_{2,c}/2}} 
\;=\; \frac{1}{2\sqrt{2}} ~,
\ea
in agreement with \cite{DSI}. For the antispinodal point one finds
\ba
\eta_a  \;=\; \frac{1}{4} ~, \hspace{1cm}
r_{1a} \;=\; 0 ~, \hspace{1cm} 
r_{2a} \;=\; \pm \frac{4\sqrt{2}}{7} ~. 
\ea
The expressions for the spinodal point are rather complicated, 
which is the reason why only numerical values are given 
\ba
\eta_s \;=\; 0.219 ~, \hspace{1cm} 
r_{1s} \;=\; 0.075 ~, \hspace{1cm} 
r_{2s} \;=\; \pm 0.748 ~.
\ea
In the right panel of Fig.~\ref{pot1} the potential-energy surface is 
given for five different values of the control parameter $\eta$. 
For $0 \leq \eta < \eta_s$ the system only has a spherical minimum. 
At $\eta=\eta_s$, it develops a second deformed minimum which becomes 
degerenate with the spherical minimum at $\eta=\eta_c$. 
The spherical minimum disappears 
at $\eta=\eta_a$, and for $\eta > \eta_a$ the system is deformed. 
For $\eta_s < \eta < \eta_a$, there is a coexistence of a spherical 
and a deformed minimum (see inset), the former is the lowest for 
$\eta_s < \eta < \eta_c$, and the latter for $\eta_c < \eta < \eta_a$. 
These features are characteristic of a first-order phase transitition. 
Recent experiments have provided strong evidence for $^{152}$Sm as a nucleus 
at the critical point of a phase transition between spherical and deformed 
shapes \cite{IZC}. 

\subsection*{The $SO(6) - SU(3)$ transitional region}

Finally, the transitional region between the $SO(6)$ and $SU(3)$ 
limits can be studied by  
\ba
H_{23} \;=\; H_2 + H_3 \;=\; - \kappa' \, \hat Q(0) \cdot \hat Q(0) 
- \kappa \, \hat Q(\chi) \cdot \hat Q(\chi) ~, \hspace{1cm} 
\chi=\mp \frac{1}{2} \sqrt{7} ~.
\ea
as a function of the control parameter $\xi=\kappa/\kappa'$ with 
$0 \leq \xi < \infty$. The corresponding energy surface is given by
\ba
{\cal E}(\beta,\gamma) &=& \frac{V_2(\beta)+V_3(\beta,\gamma)}{N(N-1)\kappa'} 
\nonumber\\
&=& - \frac{4\beta^2}{(1+\beta^2)^2} - \frac{\xi}{(1+\beta^2)^2} 
\left[ 4 \beta^2 \pm 2 \sqrt{2} \beta^3 \cos 3\gamma 
+ \frac{1}{2} \beta^4 \right] ~,
\label{v23}
\ea
where the energy functionals of Eqs.~(\ref{v2}) and~(\ref{v3}) are scaled  
by $N(N-1)\kappa'$. In this case, there is no phase transition \cite{DSI}. 

In terms of the coefficients $r_1$ and $r_2$ one finds 
\ba
r_1 \;=\; - \frac{4+4\xi}{4+3\xi} ~, \hspace{1cm}
r_2 \;=\; \pm \frac{4\xi\sqrt{2}}{4+3\xi} ~. 
\ea
Also this transition corresponds to a straight line in the $r_2$-$r_1$ 
plane 
\ba
r_1 \;=\; -1 \mp \frac{r_2}{4\sqrt{2}} ~, \hspace{1cm} 
-\frac{4}{3} \leq r_1 \leq -1 ~.
\label{trans23}
\ea
Since the line defined in Eq.~(\ref{trans23}) does not intersect the 
curves for the critical, spinodal and antispinodal points, $H_{23}$ 
does not exhibit a phase transition. 

\subsection{Shape-phase diagram}

\begin{figure}
\centerline{\epsfig{file=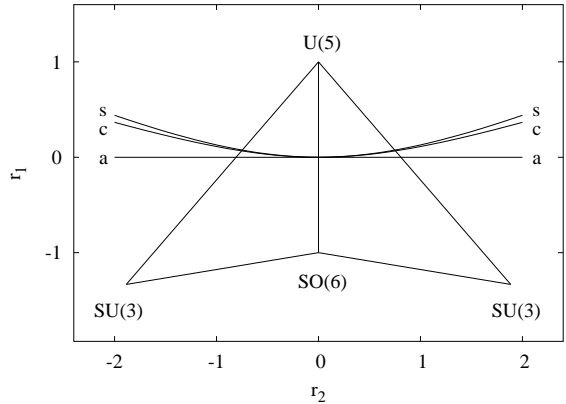,angle=270,width=0.5\textwidth}} 
\caption[Shape-phase diagram]
{\small Shape-phase diagram for the IBM as a function of $r_1$ and $r_2$. 
The curves marked by $c$, $s$ and $a$ correspond to the critical, spinodal 
and antispinodal points, respectively.}
\label{diagram}
\end{figure}

In Fig.~\ref{diagram}, the results of the previous section are summarized  
in terms of a shape-phase diagram of the IBM plotted as a function of the 
coefficients $r_1$ and $r_2$ of Eq.~(\ref{control}). These coefficients 
are determined by the parameters of the Hamiltonian and completely 
classify the corresponding geometric shapes. A given Hamiltonian 
corresponds to a point in the $r_2$-$r_1$ plane. However, note that 
the opposite is not true: a given point in the $r_2$-$r_1$ plane corresponds 
not to a single Hamiltonian, but to a class of Hamiltonians which all share 
the same potential-energy surface, but differ in their kinetic energies. 

The transitional regions between the dynamical symmetries form a kite 
(or a double triangle) spanned by the straight lines derived for each one 
of them, see Eqs.~(\ref{trans12}), (\ref{trans13}) and (\ref{trans23}). 
The dynamical symmetries occur at the end points of the kite, $U(5)$ 
(spherical), $SO(6)$ ($\gamma$-unstable deformed) and $SU(3)$ (prolate 
and oblate deformed). The critical points for the phase transition between 
spherical and deformded nuclei all lie on the curve defined by 
Eq.~(\ref{r1c}) which is labeled by ``c'' in Fig.~\ref{diagram}. 
The spinodal and antispinodal points are labeled by ``s'' and ``a'', 
respectively.  

The region above the spinodal points ($r_1>r_{1s}$) corresponds to 
spherical equilibrium shapes with only a spherical minimum. At the spinodal 
point ($r_1=r_{1s}$), the system develops a second deformed minimum which 
becomes degerenate with the spherical minimum at the critical point 
($r_1=r_{1c}$). At the antispinodal point ($r_1=r_{1a}$), the spherical 
minimum disappears and the region below the antispinodal points ($r_1<r_{1a}$) 
corresponds to deformed equilibrium shapes with only a deformed minimum. 
For $r_{1s} < r_1 < r_{1a}$, there is a coexistence of a spherical 
and a deformed minimum, the former is the lowest for 
$r_{1s} < r_1 < r_{1c}$ and the latter for $r_{1c} < r_1 < r_{1a}$. 
These are the characteristic features of a first-order phase 
transitition. On the line $r_2=0$, the spinodal, critical and antispinodal 
points coincide, corresponding to an isolated point with a second-order 
phase transition. The line $r_2 = 0$ and $r_1 < 0$ has been identified 
with a first-order quantum phase transition for $\gamma$-soft nuclei 
\cite{ELM,Jolie}.

An often-used representation of the phase space of 
the IBM is the so-called symmetry or Casten triangle which was introduced 
in \cite{Rick} as a schematic way to represent the properties of the 
Hamiltonian: the dynamical symmetries at the corners, the transitions 
between two of them along the sides, and arbitrary combinations of the 
three of them in the inside of the triangle \cite{QPT}. 
In Section~\ref{shapes} it 
was shown that the catastrophe theory treatment of \cite{ELM} provides a 
derivation of this triangle, not as a schematic representation, but as 
a function of two control parameters $r_1$ and $r_2$ which depend on the 
parameters of the Hamiltonian. This method is valid not only for the 
schematic Hamiltonian discussed in this contribution, but also for arbitrary 
IBM Hamiltonians.

\subsection{Shape-phase coexistence}

For first-order phase transitions between spherical and deformed shapes 
there is a very small region of coexistence. The right panel of 
Fig.~\ref{pot1} shows that the barrier that separates the spherical and 
deformed minima is very small, even in the inset it is hard to see. 
The height of the barrier is only a few percent of the excitation energy 
of the $2^+_1$ state. In a recent study, evidence has been presented for 
the existence of such coexisting phases at low energy in the transitional 
nucleus $^{152}$Sm \cite{IZC}.

However, it is important to distinguish between this type of shape 
coexistence which occurs for spherical-deformed transitional nuclei and 
the far more common phenomenon of coexisting shapes due to different 
configurations \cite{coex,kris}. The latter situations have been treated in an 
extension of the IBM in terms of a configuration mixing calculation, 
in which the normal and intruder configurations are described by IBM 
Hamiltonians and the configuration mixing by a coupling term \cite{phil} 
\ba
\left( \begin{array}{cc} H_{\rm normal} & V_{\rm mix} \\
V_{\rm mix} & H_{\rm intruder} \end{array} \right) ~. 
\ea

\begin{figure}
\vfill 
\begin{minipage}{.5\linewidth}
\centerline{\epsfig{file=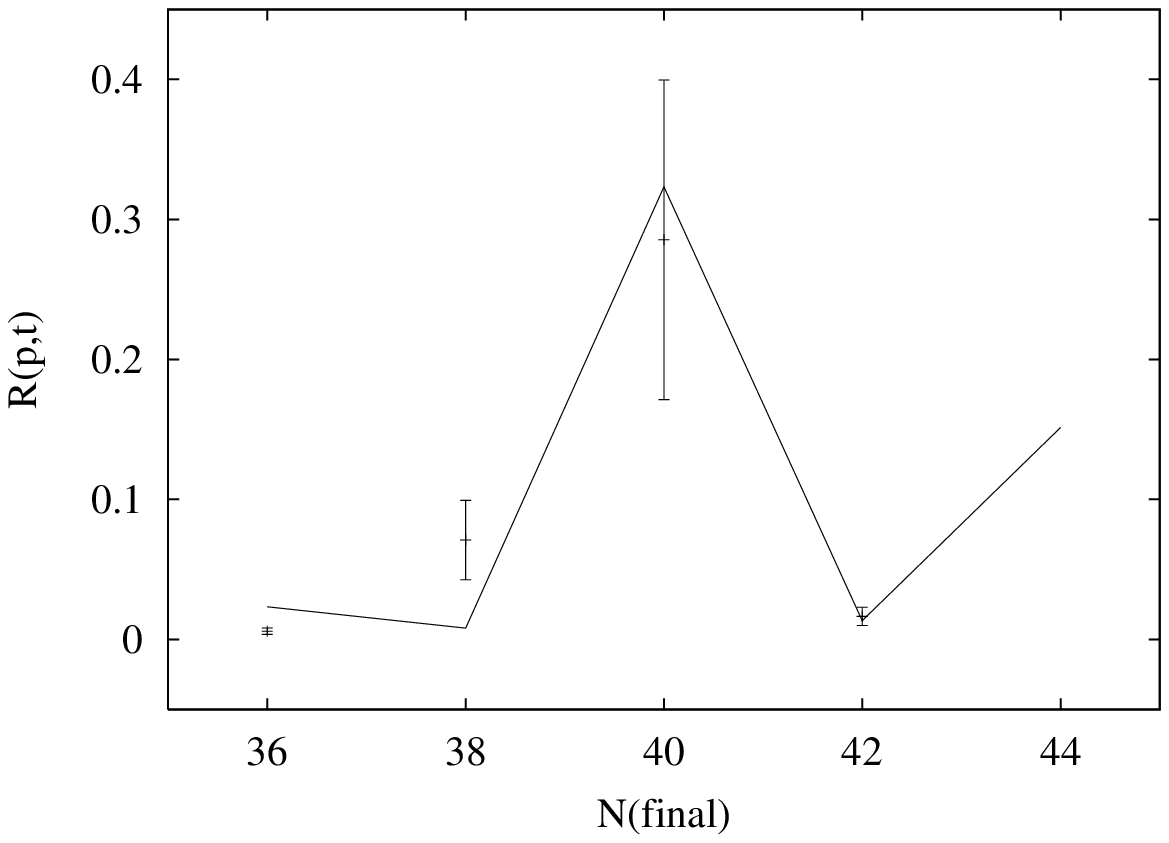,width=\linewidth}}
\end{minipage}\hfill
\begin{minipage}{.5\linewidth}
\centerline{\epsfig{file=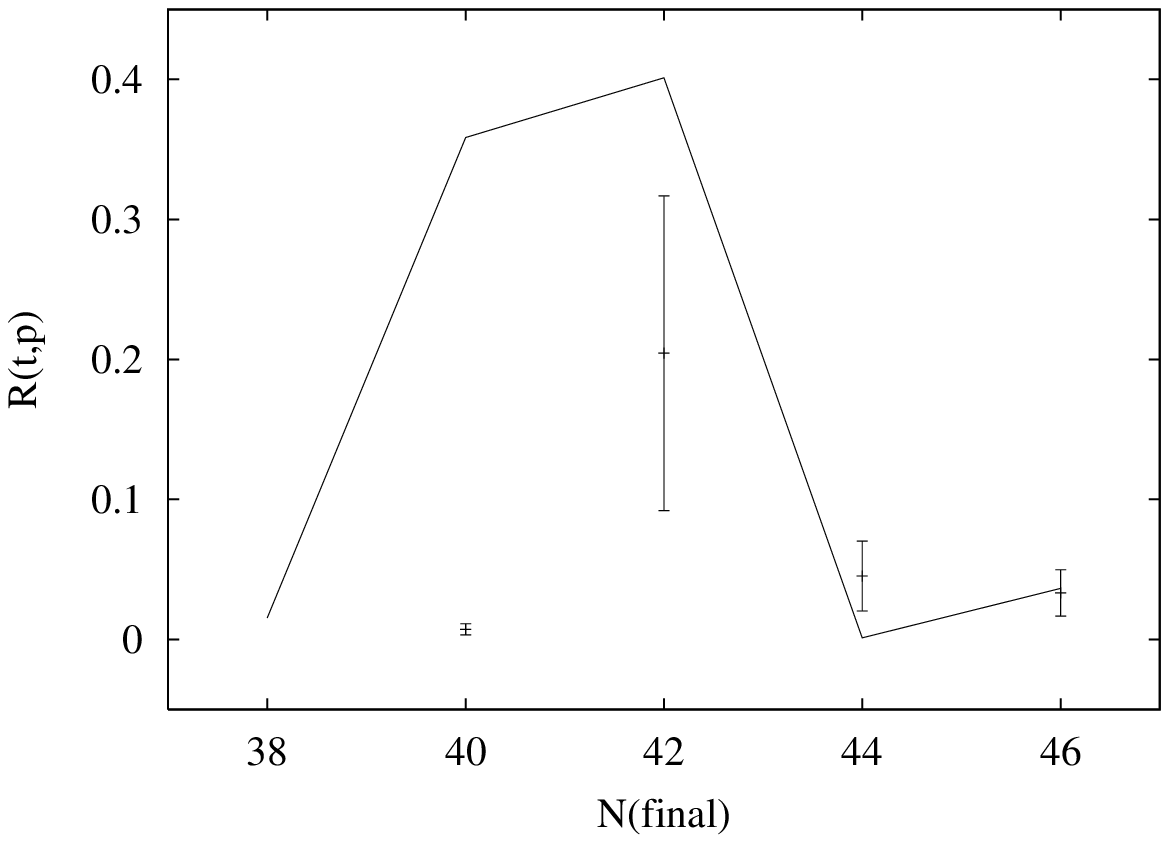,width=\linewidth}}
\end{minipage}
\caption[Two-neutron transfer reactions]
{\small Comparison between experimental and theoretical cross sections for 
two-neutron transfer reactions $(p,t)$ and $(t,p)$ between Ge isotopes. 
$N$ denotes the number of neutrons in the final Ge nucleus.}
\label{rpt}
\end{figure}

In this subsection, I briefly discuss some recent results obtained for the 
Ge isotopes \cite{duval,EPR}. The even-even Ge isotopes show strong evidence 
for the coexistence of different geometric shapes. A sensitive probe for the 
configuration mixing is provided by two-neutron transfer reactions $(p,t)$ 
and $(t,p)$, in particular the ratio of cross sections for the excitation 
of the $0^+_2$ state and the ground state
\ba
R \;=\; \frac{\sigma(0^+_2 \rightarrow 0^+_1)}
             {\sigma(0^+_1 \rightarrow 0^+_1)} ~.
\ea
In the absence of mixing, this ratio is zero ($U(5)$ limit) or very small 
($SU(3)$ and $SO(6)$ limits) \cite{AI}. However, if the two configurations are 
strongly mixed, $R$ can be different from zero. In Fig.~\ref{rpt}, I show 
the results for $(p,t)$ and $(t,p)$ reactions between the Ge isotopes 
\cite{EPR}. The enhancement in the cross sections around $N = 40$ is an 
indictation of the importance of configuration mixing.

\subsection{Triaxial degrees of freedom}

The one- and two-body IBM-1 Hamiltonian can give rise to spherical, 
axially deformed and $\gamma$-unstable deformed shapes. 
There is no stable triaxially deformed shape, unless one includes 
three-body interactions \cite{Piet}. In the IBM-2 in which one 
distinguishes between protons and neutrons, there is in addition 
to the $U(5)$, $SU(3$ and $SO(6)$ limits, another dynamical symmetry 
called the $SU^{\ast}(3)$ limit, which has been associated with 
triaxial shapes \cite{su3star}. It is therefore of great interest to 
extend the discussion of phase transitions to IBM-2, since there are 
new transitional regions which are not present in IBM-1, namely 
$U(5)-SU^{\ast}(3)$, $SU(3)-SU^{\ast}(3)$ and $SO(6)-SU^{\ast}(3)$. 
Let's consider a schematic IBM-2 Hamiltonian
\ba
H \;=\; \epsilon \, \left( n_{d_{\pi}} + n_{d_{\nu}} \right) 
-\kappa \, \left[ Q_{\pi}(\chi_{\pi}) + Q_{\nu}(\chi_{\nu}) \right] 
\cdot \left[ Q_{\pi}(\chi_{\pi}) + Q_{\nu}(\chi_{\nu}) \right] ~,
\label{hibm2}
\ea
consisting of the $d$-boson energies and a quadrupole-quadrupole 
interaction. The $SU^{\ast}(3)$ limit arises for $\epsilon=0$ and 
$\chi_{\pi}=-\chi_{\nu}=\pm \sqrt{7}/2$. 

The analysis of the equilibrium shapes becomes a lot more complicated 
since in general one now has to study the properties of an energy 
surface that depends on 7 geometric variables: $\beta_{\pi}$, 
$\gamma_{\pi}$, ($\beta_{\nu}$, $\gamma_{\nu}$) for the geometry of 
the proton (neutron) distribution and three Euler angles 
$(\Theta) \equiv (\Theta_1,\Theta_2,\Theta_3)$ for the relative orientation 
of the proton and the neutron distributions. The general form of the 
potential-energy surface of IBM-2 can be found in \cite{Ami}. 
In the same reference it was shown that only IBM-2 Hamiltonians 
with a repulsive hexadecupole-hexadecupole interaction can give rise 
to so-called oblique shapes, {\it i.e.} an equilibrium shape with 
$\Theta \neq 0$. Therefore, for the Hamiltonian of Eq.~(\ref{hibm2}) one 
can take $\Theta=0$ \cite{Arias,Caprio}. The energy surface is then given 
by the simplified form 
\ba
V(\beta_{\pi},\gamma_{\pi},\beta_{\nu},\gamma_{\nu}) &=& 
V(\beta_{\pi},\gamma_{\pi}) + V(\beta_{\nu},\gamma_{\nu}) 
\nonumber\\
&& \left. - \frac{2\kappa N_{\pi} N_{\nu} \beta_{\pi} \beta_{\nu}} 
{(1+\beta^2_{\pi})(1+\beta^2_{\nu})} 
\right[ 4 \cos(\gamma_{\pi}-\gamma_{\nu}) 
- 2 \sqrt{\frac{2}{7}} \, \chi_{\nu} \beta_{\nu} 
\cos(\gamma_{\pi}+2\gamma_{\nu}) 
\nonumber\\
&& \left. - 2 \sqrt{\frac{2}{7}} \, \chi_{\pi} \beta_{\pi} 
\cos(\gamma_{\nu}+2\gamma_{\pi}) 
+ \frac{2}{7} \, \chi_{\pi} \chi_{\nu} \beta_{\pi} \beta_{\nu} 
\cos(2\gamma_{\pi}-2\gamma_{\nu}) \right] ~, 
\ea
where $V(\beta_{\rho},\gamma_{\rho})$ is the energy surface for the 
proton ($\rho=\pi$) and the neutron ($\rho=\nu$) part of the Hamiltonian, 
see Eq.~(\ref{eibm1}). 

The results that have been obtained so far for IBM-2 are mostly numerical 
\cite{Arias,Caprio}: there is a second-order phase transition for 
$U(5)-SU^{\ast}(3)$ and $SU(3)-SU^{\ast}(3)$, whereas for 
$SO(6)-SU^{\ast}(3)$ there is no phase
transition. It would be very interesting to have analytic results for the 
phase transitions between the four dynamical symmetries of the 
IBM-2. In this respect, the methods of catastrophe theory could be of 
help to obtain closed expressions for the critical and (anti)spinodal 
points, to provide insight into the order of the new phase transtions, 
and perhaps even to derive a shape-phase diagram for IBM-2, 
much in the same way as was done for the IBM-1 \cite{ELM}. 

\section{Random interactions}

A different method to study the geometric shapes of the IBM, and especially 
their robustness, is that of Hamiltonians with random interactions. 
   
Random matrix theory was developed to describe statistical 
properties of nuclear spectra, such as average distributions and 
fluctuations of peaks in neutron-capture experiments 
\cite{Wigner,Porter}. In this approach, the Hamiltonian matrix 
elements are chosen at random, while keeping some global symmetries, 
{\it i.e.} the matrix should be hermitean, and be invariant under 
time-reversal, rotations and reflections. Specific examples include 
the Gaussian Orthogonal Ensemble (GOE) of real-symmetric random 
Hamiltonian matrices in which the many-body interactions are 
uncorrelated, and the Two-Body Random Ensemble (TBRE) in which the 
two-body interactions are taken from a distribution of random 
numbers \cite{Brody}. For two-particle systems the two ensembles are 
identical but for more than two particles, unlike in the case of GOE, 
in TBRE the many-body matrix elements are correlated. As a 
consequence, also the energy eigenvalues of states with different 
quantum numbers are strongly correlated, since they arise from the 
same Hamiltonian. 

The latter aspect was investigated in shell model 
calculations for even-even nuclei in the $sd$ shell and 
the $pf$ shell \cite{JBD}. An analysis of the energy spectra of an 
ensemble of random two-body Hamiltonians showed a remarkable 
statistical preference for ground states with angular momentum and 
parity $L^P=0^+$, despite the random nature of the two-body matrix 
elements both in sign and relative magnitude. A similar 
preponderance of $0^+$ ground states was found in an analysis 
of the IBM with random interactions 
\cite{BF1}. In addition, in the IBM evidence was found for both 
vibrational and rotational band structures. According to the 
conventional ideas in the field, the occurrence of $L=0$ ground states 
and the existence of vibrational and rotational bands are due to  
very specific forms of the interactions. Therefore, these unexpected 
and surprising results with random interactions have sparked a large 
number of investigations to explain and further explore the properties 
of random nuclei (see {\it e.g.} the reviews by Bijker and Frank 
\cite{BF5}, Zelevinsky and Volya \cite{Zelevinsky} and Zhao, Arima and 
Yoshinaga \cite{Zhao}).  

In this section, I discuss the phenomenon of emerging regular spectral 
features from the IBM with random interactions and its 
relation with the underlying geometric shapes and critical points. 

\subsection{Regular features and robust results}

In order to study the geometric shapes associated with the IBM, let's 
consider the most general one- and two-body IBM Hamiltonian
\ba
H \;=\; \frac{1}{N} \left[ H_{1,b} + \frac{1}{N-1} H_{2,b} \right] ~,
\label{hibm}
\ea
where $H_{1,b}$ contains the boson energies
\ba
H_{1,b} \;=\; \epsilon_s \, s^{\dagger} \cdot \tilde{s}  
+ \epsilon_d \, d^{\dagger} \cdot \tilde{d} ~,
\label{h1ibm}
\ea
and $H_{2,b}$ the two-body interactions
\ba
H_{2,b} &=& u_0 \, \frac{1}{2} \, (s^{\dagger} \times s^{\dagger})^{(0)} \cdot 
(\tilde{s} \times \tilde{s})^{(0)} 
+ u_2 \, (s^{\dagger} \times d^{\dagger})^{(2)} \cdot 
(\tilde{d} \times \tilde{s})^{(2)} 
\nonumber\\ 
&& + \sum_{\lambda=0,2,4} c_{\lambda} \, \frac{1}{2} \, 
(d^{\dagger} \times d^{\dagger})^{(\lambda)} \cdot  
(\tilde{d} \times \tilde{d})^{(\lambda)} 
\nonumber\\
&& + v_0 \, \frac{1}{2\sqrt{2}} \, \left[ 
  (d^{\dagger} \times d^{\dagger})^{(0)} \cdot 
  (\tilde{s} \times \tilde{s})^{(0)} 
+ (s^{\dagger} \times s^{\dagger})^{(0)} \cdot 
  (\tilde{d} \times \tilde{d})^{(0)} \right]
\nonumber\\
&& + v_2 \, \frac{1}{2} \, \left[ 
  (d^{\dagger} \times d^{\dagger})^{(2)} \cdot 
  (\tilde{d} \times \tilde{s})^{(2)} 
+ (s^{\dagger} \times d^{\dagger})^{(2)} \cdot 
  (\tilde{d} \times \tilde{d})^{(2)} \right] ~.
\label{h2ibm}
\ea
The nine parameters of the IBM Hamiltonian 
of Eqs.~(\ref{hibm})-(\ref{h2ibm}), altogether denoted by
\ba
(\vec{x}) \;\equiv\; (\epsilon_s, \epsilon_d, 
u_0, u_2, c_0, c_2, c_4, v_0, v_2) ~,
\ea
are taken as independent random numbers $x_i$ ($i=1,\ldots,9$) on a 
Gaussian distribution 
\ba
P(x_i) &=& \mbox{e}^{-x_i^2/2\sigma^2}/\sigma\sqrt{2\pi} ~, 
\label{gauss}
\ea
with zero mean and width $\sigma$. In this way, the interaction terms 
are arbitrary and equally likely to be attractive or repulsive. The 
spectral properties of each Hamiltonian are then analyzed by exact numerical 
diagonalization \cite{BF1}. 

\begin{figure}[b]
\vfill 
\begin{minipage}{.5\linewidth}
\centerline{\epsfig{file=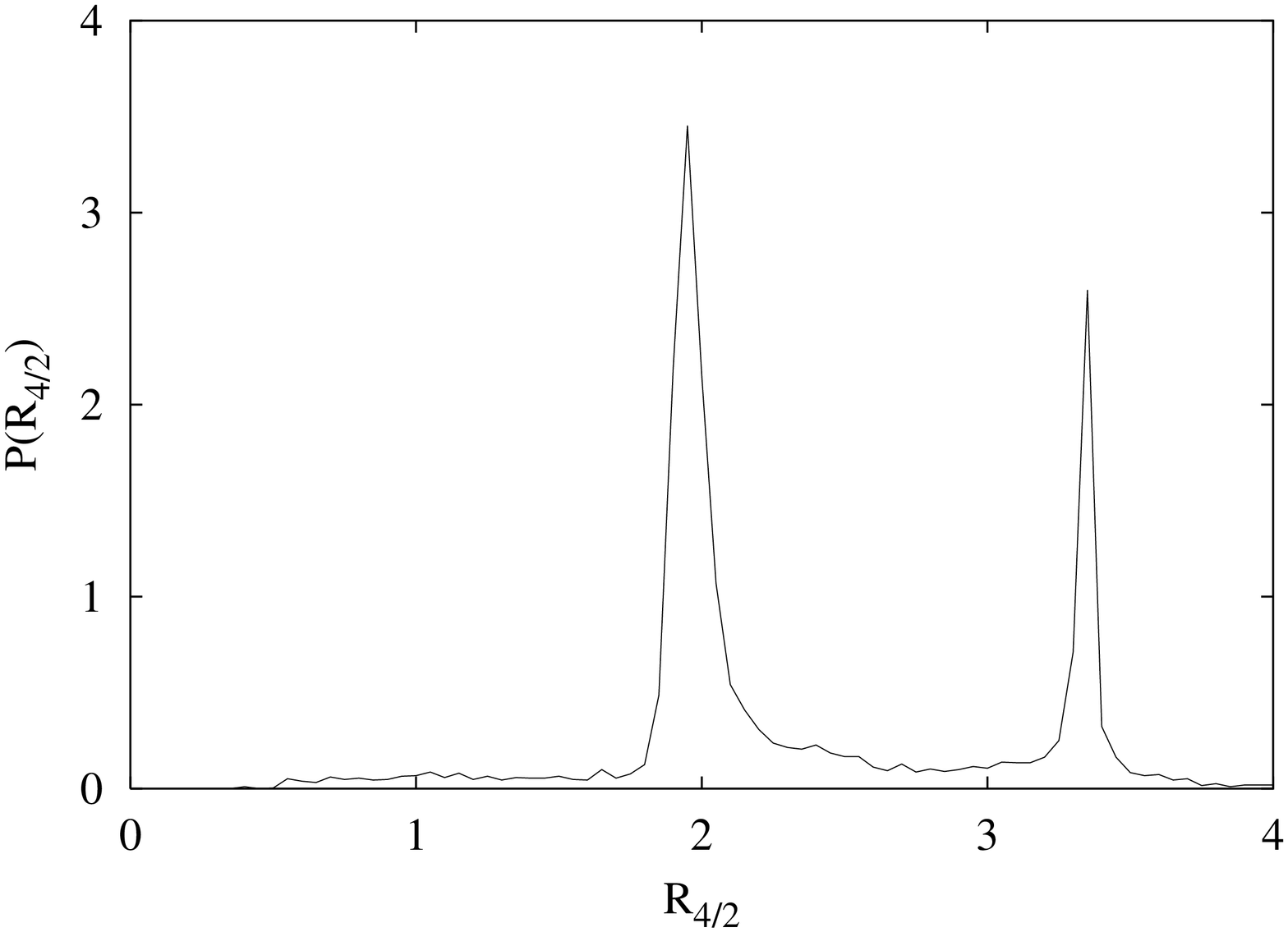,angle=270,width=\linewidth}}
\end{minipage}\hfill
\begin{minipage}{.5\linewidth}
\centerline{\epsfig{file=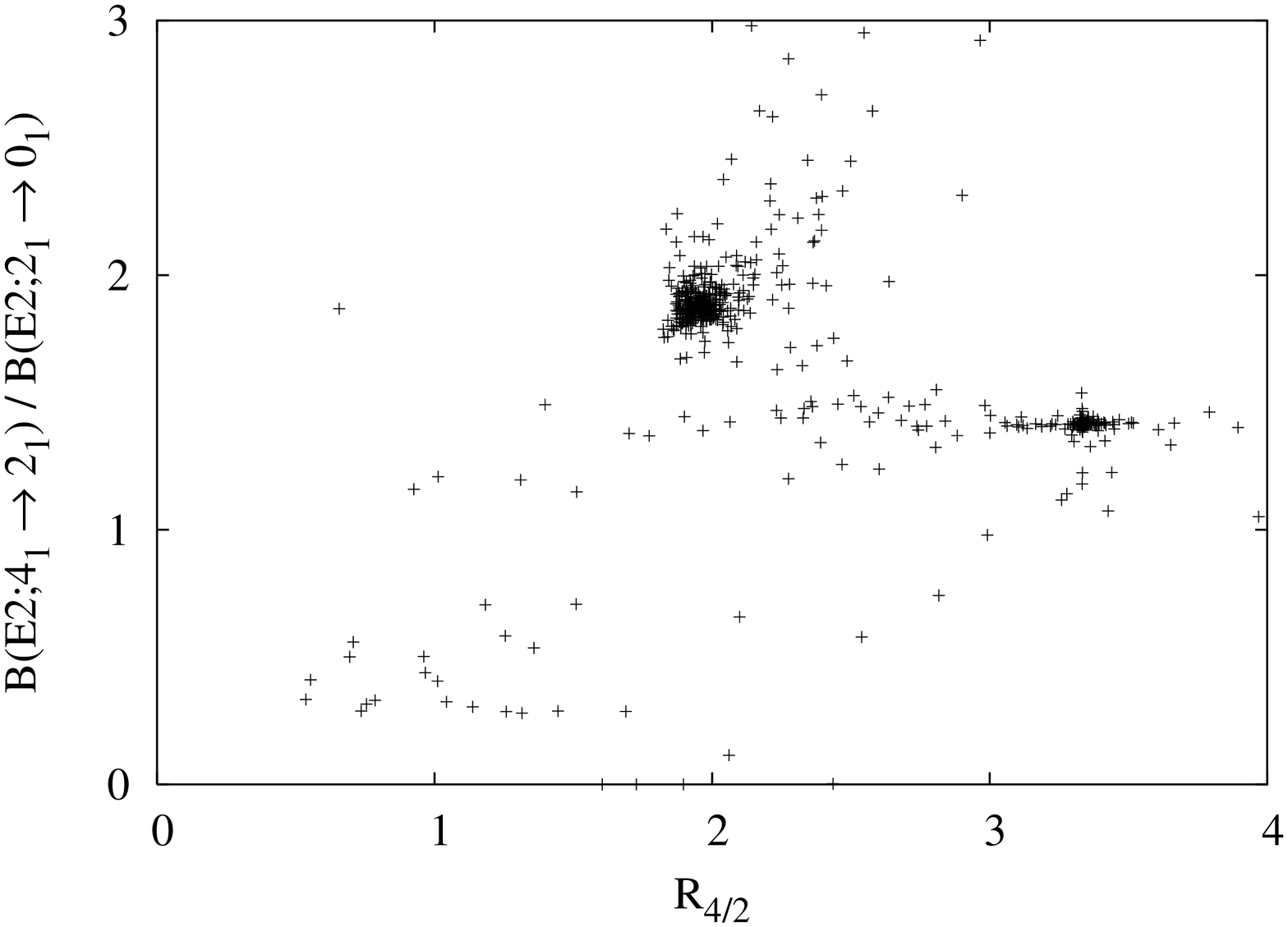,angle=270,width=\linewidth}}
\end{minipage}
\caption[Energy ratios for CQF]
{\small Probability distribution $P(R_{4/2})$ of the energy ratio $R_{4/2}$ 
(left) and correlation between ratios of $B(E2)$ values and energies (right) 
for the IBM with random one- and two-body interactions. 
The number of bosons is $N = 16$.}
\label{ratio}
\end{figure}

The properties of the energy spectra can be studied by the energy ratio
\ba
R_{4/2} \;=\; \frac{E(4^+_1)-E(0^+_1)}{E(2^+_1)-E(0^+_1)} ~. 
\label{r42}
\ea
In Table~\ref{BE2}, the values of $R_{4/2}$ are presented for each of the 
dynamical symmetries: $R_{4/2} = 2.0$, $2.5$ and $3.3$ for the $U(5)$, $SO(6)$ 
and $SU(3)$ limits, respectively. Fig.~\ref{ratio} shows a remarkable result: 
the probability distribution $P(R_{4/2})$ has two very pronounced peaks, 
one at $\sim 1.95$ and a narrower one at $\sim 3.35$. These values 
correspond almost exactly to the harmonic vibrator and rotor values 
(see the results for the $U(5)$ and $SU(3)$ limits in Table~\ref{BE2}).  
No such peak is observed for the $\gamma$-unstable or deformed oscillator 
case ($SO(6)$ limit). This can be understood from the fact that $\gamma$ 
dependence of the potential-energy surface arises from a single parameter, 
$v_2$. For a $\gamma$-unstable solution to occur, $v_2$ has to be zero. 
For $v_2 \ne 0$ the energy surface has a prolate or an oblate minimum 
(whether the global minimum is spherical or deformed depends on the other 
terms in the Hamiltonian as well).

\begin{table} 
\centering
\caption[Energies and B(E2) values]
{\small Energies of $B(E2)$ values in the dynamical symmetry 
limits of the IBM for the Hamiltonians of Eqs.~(\ref{h1}), (\ref{h2}) 
and (\ref{h3}).}
\label{BE2}
\vspace{15pt}
\begin{tabular}{cccccc}
\hline
& & & & & \\
& $\frac{E(4^+_1)-E(0^+_1)}{E(2^+_1)-E(0^+_1)}$  
& $\frac{E(0^+_2)-E(0^+_1)}{E(2^+_1)-E(0^+_1)}$  
& \multicolumn{3}{c}{$\frac{B(E2;4^+_1 \rightarrow 2^+_1)}
{B(E2;2^+_1 \rightarrow 0^+_1)}$} \\
& & & & & \\
\hline
& & & & & \\
$U(5)$ & $2$ & $2$ & $\frac{2(N-1)}{N}$ & $\rightarrow$ & $2$ \\
& & & & & \\
$SO(6)$ & $\frac{5}{2}$ & $N+1$  
& $\frac{10(N-1)(N+5)}{7N(N+4)}$ & $\rightarrow$ & $\frac{10}{7}$ \\
& & & & & \\
$SU(3)$ & $\frac{10}{3}$ & $\frac{4(2N-1)}{3}$ 
& $\frac{10(N-1)(2N+5)}{7N(2N+3)}$ & $\rightarrow$ & $\frac{10}{7}$ \\
& & & & & \\
\hline
\end{tabular}
\end{table}

Energies by themselves are not sufficient to decide whether 
or not there exists band structure. Levels belonging to a 
collective band are connected by strong electromagnetic 
transitions. In Fig.~\ref{ratio} a correlation plot is shown between 
the ratio of $B(E2)$ values for the $4^+_1 \rightarrow 2^+_1$ and 
$2^+_1 \rightarrow 0^+_1$ transitions and the energy ratio $R_{4/2}$. 
Table~\ref{BE2} shows that in the large $N$ limit 
this ratio of $B(E2)$ values is $2$ for the harmonic oscillator 
($U(5)$ limit) and $10/7$ for the deformed oscillator and the rotor 
($SO(6)$ and $SU(3)$ limit, respectively). 
The right panel of Fig.~\ref{ratio} shows a strong correlation 
between the first peak in the energy ratio and the vibrator value 
for the ratio of $B(E2)$ values (the concentration of points in this 
region corresponds to about 50 $\%$ of all cases), as well as for 
the second peak and the rotor value (about 25 $\%$ of all cases).  

\begin{figure}
\centerline{\epsfig{file=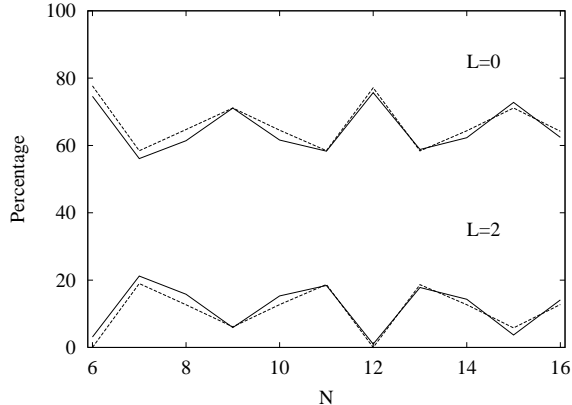,width=0.5\textwidth}}
\caption[Ground state angular momentum]
{\small Percentages of ground states with $L=0$ (top) and $L=2$ 
(bottom) in the IBM with random one- and two-body interactions as a 
function of the number of bosons $N$: calculated 
exactly for 10,000 runs (solid line) and in mean-field approximation 
(dashed line).}
\label{ibmgs}
\end{figure}

Despite the random nature of the interaction strengths both in relative 
size and sign, the ground state still has $L=0$ in the majority of the cases. 
Fig.~\ref{ibmgs} shows the percentages of ground states with $L=0$ 
and $L=2$ as a function of the boson number $N$ (solid line). 
One sees a clear dominance of ground states with $L=0$ with $\sim$ 60-75$\%$. 
For $N=3k$ (a multiple of 3) one sees an enhancement for $L=0$ 
and a decrease for $L=2$. The sum of the two hardly depends on the 
number of bosons. 

\subsection{Mean-field analysis}

The results obtained in studies of the nuclear shell model and the IBM 
with random interactions are very surprising and unexpected in the sense 
that, according to the conventional ideas in the field, the occurrence of 
$L=0$ ground states and the existence of vibrational and rotational bands 
are due to very specific forms of the interactions. 

The basic ingredients of the numerical simulations are the structure of the 
model space, the ensemble of random Hamiltonians, the order of the 
interactions (one- and two-body), and the global symmetries, {\it i.e.} 
time-reversal, hermiticity and rotation and reflection symmetry. 
The latter three symmetries cannot be modified, since nuclear levels have 
real energies and good angular momentum and parity. It has been shown 
that the observed spectral order is a robust property that does 
not depend on the specific choice of the ensemble of 
random interactions \cite{JBD,BFP1,JBDT}, the time-reversal 
symmetry \cite{BFP1}, or the restriction of the Hamiltonian to one- 
and two-body interactions \cite{BF2}. These results suggest that 
that an explanation of the origin of the observed regular features 
has to be sought in the many-body dynamics of the model space 
and/or the general statistical properties of random interactions. 

In this section, the origin of the regular features that 
emerge from random interactions in the IBM \cite{BF3} are studied 
using standard Hartree-Bose mean-field methods \cite{duke} 
in which the trial wave function is the boson condensate of Eqs.~(\ref{cost}) 
and (\ref{condensate}) (see Section~\ref{pes}). 

\subsection*{Vibrations} 

The potential-energy surface associated with the IBM Hamiltonian of 
Eqs.~(\ref{hibm})-(\ref{h2ibm}) is given by 
its expectation value in the coherent state 
\ba
V(\beta,\gamma) &=& a_0 + a_2 \frac{\beta^2}{1+\beta^2} 
\frac{1}{(1+\beta^2)^2} \left[ a_3 \beta^3 \cos 3\gamma 
+ a_4 \beta^4 \right] ~, 
\label{vbg}
\ea
where the coefficients $a_i$ are linear combinations of the parameters 
of the Hamiltonian 
\ba
a_4 &=& \frac{1}{2} u_0 - u_2 
+ \frac{1}{10} c_0 + \frac{1}{7} c_2 + \frac{9}{35} c_4 
- \frac{1}{\sqrt{10}} v_0 ~, 
\nonumber\\
a_3 &=& -\sqrt{\frac{2}{7}} v_2 ~,
\nonumber\\
a_2 &=& \epsilon_d-\epsilon_s - u_0 + u_2 
+ \frac{1}{\sqrt{10}} v_0 ~, 
\nonumber\\
a_0 &=& \epsilon_s + \frac{1}{2} u_0 ~. 
\label{coef}
\ea
The energy surface of Eq.~(\ref{vbg}) provides information on the 
distribution of shapes that the model can acquire. The values of 
$\beta_e$ and $\gamma_e$ that characterize the equilibrium 
configuration of the potential-energy surface depend on the 
coefficients $a_4$, $a_3$ and $a_2$. 
The equilibrium shapes can be divided into three different classes:
\begin{itemize}
\item $\beta_e = 0$: an $s$-boson or spherical condensate 
\item $0 < \beta_e < \infty$ with $\gamma = 0^{\circ}$ or 
$\gamma = 60^{\circ}$: a deformed condensate with prolate or oblate 
symmetry, respectively, and 
\item $\beta_e=\infty$: a $d$-boson condensate. 
\end{itemize} 

\subsection*{Rotations}

Each equilibrium configuration has its own characteristic angular momentum 
content. Even though the angular momentum states are not projected from the 
coherent state, the ground state angular momentum  
can be obtained from the rotational structure of the condensate in 
combination with the Thouless-Valatin formula for the corresponding moments 
of inertia. This procedure is described in detail in \cite{duke}. 
The application to the IBM with random interactions leads to the 
following results.
\begin{itemize}

\item The $s$-boson condensate corresponds to a spherical shape. 
Whenever such a condensate occurs (in 39.4 $\%$ of the cases), 
the ground state has $L=0$.

\item The deformed condensate corresponds to an axially symmetric deformed 
rotor. The ordering of the rotational energy levels $L=0,2,\ldots,2N$ is 
determined by the sign of the moment of inertia 
\ba
E_{\rm rot} \;=\; \frac{1}{2{\cal I}_3} L(L+1) ~. 
\ea
The moment of inertia ${\cal I}_3$ depends in a complicated way on the 
parameters in the Hamiltonian. 
The deformed condensate occurs in 36.8 $\%$ of the cases. For ${\cal I}_3>0$ 
the ground state has $L=0$ (23.7 $\%$), while for ${\cal I}_3<0$ the ground 
state has the maximum value of the angular momentum $L=2N$ (13.1 $\%$). 

\item The $d$-boson condensate corresponds to a quadrupole oscillator with 
$N$ quanta. Its rotational structure has a more complicated structure, 
since the rotational excitation energies depend on two moments of inertia 
\ba
E_{\rm rot} \;=\; \frac{1}{2{\cal I}_5} \tau(\tau+3) 
+ \left(\frac{1}{2{\cal I}_3} - \frac{1}{6{\cal I}_5} \right) L(L+1) ~, 
\ea
which are associated with the spontaneously broken three- and 
five-dimensional rotational symmetries of the $d$-boson condensate. 
The $d$-boson condensate occurs in the remaining 23.8 $\%$ of the cases. 
The results in Table~\ref{percibm} can be understood qualitatively as 
follows. For ${\cal I}_5>0$ the ground state has $\tau=0$ 
for $N$ even or $\tau=1$ for $N$ odd ($\sim$ 4 $\%$), while for 
${\cal I}_5<0$ the ground state has the maximum value of the boson 
seniority $\tau=N$ ($\sim$ 19 $\%$). For $\tau=0$ and $\tau=1$ 
there is a single angular momentum state with $L=0$ and $L=2$, 
respectively. For  the $\tau=N$ multiplet, the angular momentum of the 
ground state depends on the sign of the moment of inertia ${\cal I}_3$. 
For ${\cal I}_3>0$ the ground state has $L=0$ for $N=3k$ or 
$L=2$ for $N \neq 3k$ (9 $\%$), while for ${\cal I}_3<0$ the ground 
state has the maximum value of the angular momentum $L=2N$ (10 $\%$). 

\end{itemize}

\begin{table}[b]
\centering
\caption[]
{\small Percentages of ground states with $L=0$, $2$ and $2N$, 
obtained in a mean-field analysis of the IBM.}
\label{percibm}
\vspace{15pt}
\begin{tabular}{crrrl}
\hline
& & & & \\
Shape & $L=0$ & $L=2$ & $L=2N$ & \\
& & & & \\
\hline
& & & & \\
$\beta_e=0$ & 39.4 $\,\%$ &  0.0 $\,\%$ &  0.0 $\,\%$ & \\
& & & & \\
$0 < \beta_e < \infty$ & 23.7 $\,\%$ &  0.0 $\,\%$ & 13.1 $\,\%$ & \\
& & & & \\
$\beta_e=\infty$ & 13.5 $\,\%$ &  0.0 $\,\%$ & 10.3 $\,\%$ & $N=6k$ \\
&  0.2 $\,\%$ & 13.2 $\,\%$ & 10.4 $\,\%$ & $N=6k+1,6k+5$ \\
&  4.4 $\,\%$ &  9.0 $\,\%$ & 10.4 $\,\%$ & $N=6k+2,6k+4$ \\
&  9.3 $\,\%$ &  4.0 $\,\%$ & 10.5 $\,\%$ & $N=6k+3$ \\
& & & & \\
\hline
\end{tabular}
\end{table}

Table~\ref{percibm} shows that the spherical and deformed 
condensates contribute constant amounts of 39.4 $\%$ and 23.7 $\%$, 
respectively, to the $L=0$ ground state percentage, whereas the contribution 
from the $d$-boson condensate depends on the number of bosons $N$. The 
$L=2$ ground states arise completely from the $d$-boson condensate solution. 
Fig.~\ref{ibmgs} shows the percentages of ground states with 
$L=0$ and $L=2$ as a function of the total number of bosons $N$. 
A comparison of the results of the mean-field analysis (dashed lines) 
and the exact ones (solid lines) shows an excellent agreement.  
There is a dominance of ground states with $L=0$ for $\sim$ 63-77 
$\%$ of the cases. Both for $L=0$ and $L=2$ there are large oscillations 
with $N$ which are entirely due to the contribution of the $d$-boson 
condensate. For $N=3k$ one sees an enhancement for $L=0$ and a 
corresponding decrease for $L=2$. In the mean-field analysis, the sum of the 
two which accounts for $\sim$ 77 $\%$ of the cases, hardly depends on the 
number of bosons, in agreement with the exact results. 
For the remaining $\sim$ 23 $\%$ of the cases, 
the ground state has the maximum value of the angular momentum $L=2N$. 

In conclusion, the emergence of regular features from the IBM with random 
interactions can be explained in a Hartree-Bose mean-field analysis of the 
random ensemble of Hamiltonians, in which different regions of the parameter 
space are associated with particular intrinsic states, which in turn 
correspond to definite geometric shapes \cite{BF3}. There are three 
solutions: a spherical shape  
carried by a single state with $L=0$, a deformed shape which corresponds 
to a rotational band with $L=0,2,\ldots,2N$, and a condensate of quadrupole 
bosons which has a more complicated angular momentum content. 
The ordering of rotational energy levels depends on the sign 
of the corresponding moments of inertia.  
The mean-field analysis explains both the distribution of ground 
state angular momenta and the occurrence of vibrational and rotational 
bands. The same conclusions hold for the vibron model for which a large 
part of the results has been obtained analytically \cite{BF4}. 

\subsection{Evolution of quasi-beta energies}

\begin{figure}[b]
\vfill 
\begin{minipage}{.5\linewidth}
\centerline{\epsfig{file=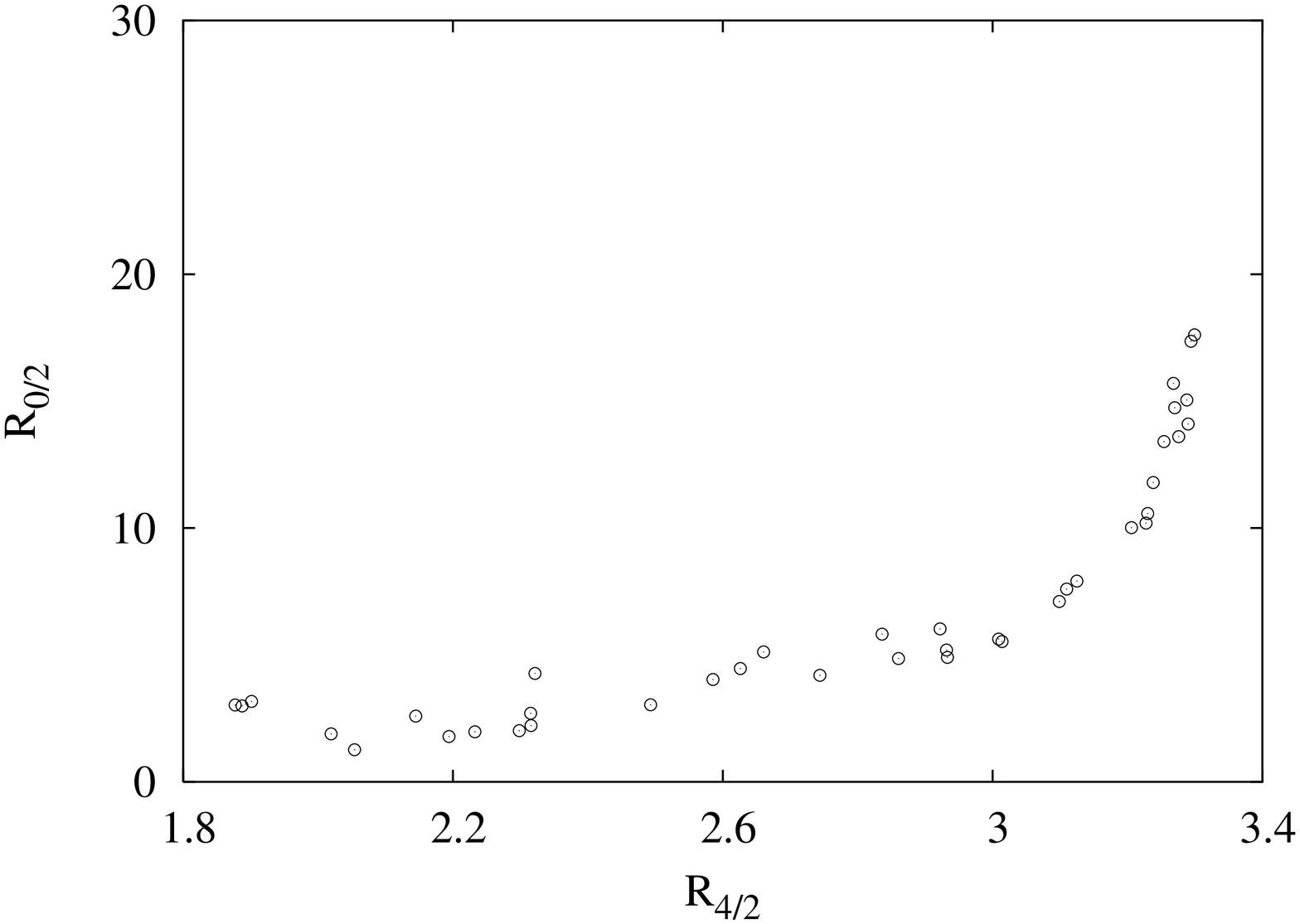,angle=270,width=\linewidth}}
\end{minipage}\hfill
\begin{minipage}{.5\linewidth}
\centerline{\epsfig{file=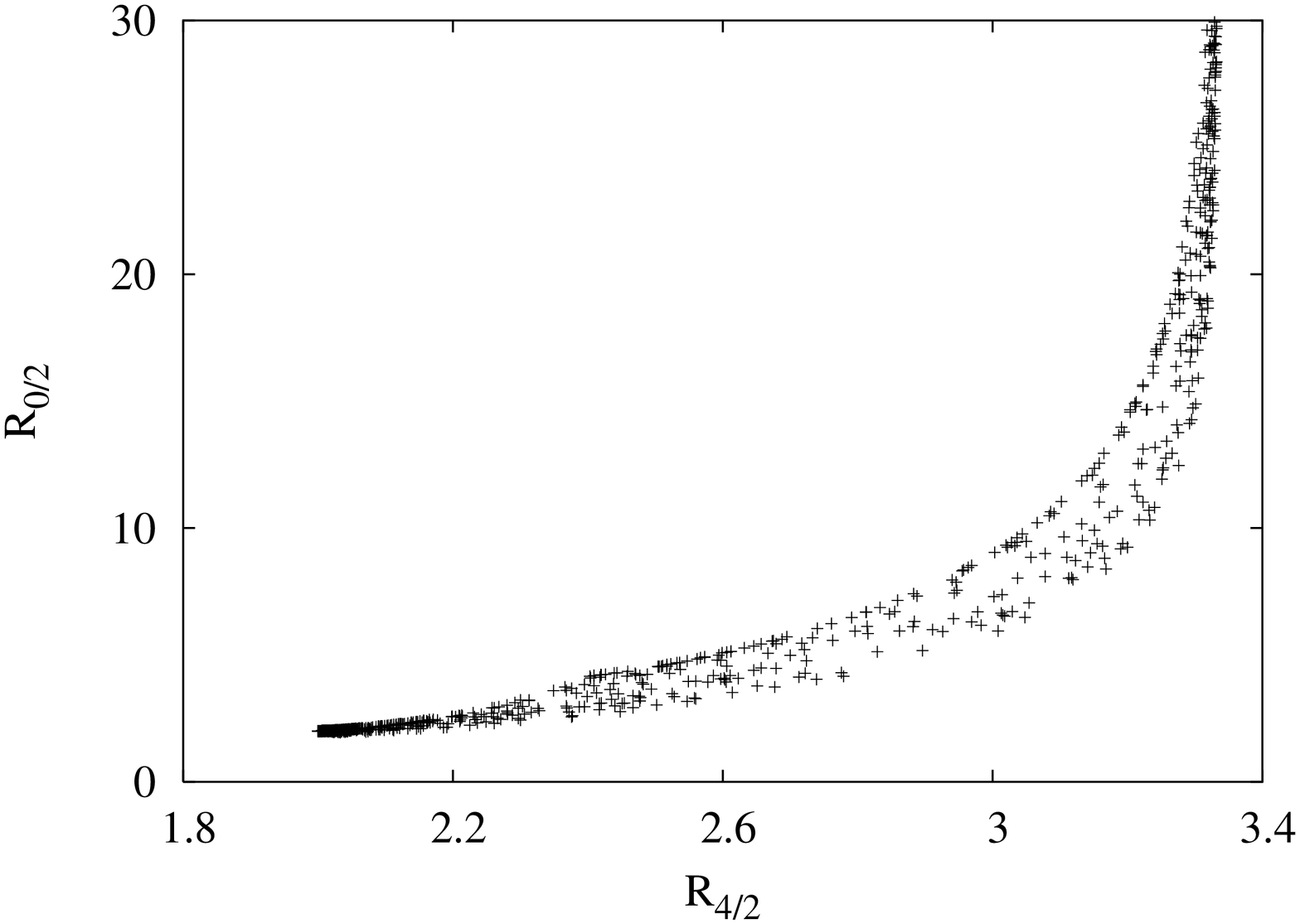,angle=270,width=\linewidth}}
\end{minipage}
\caption[Quasi-beta systematics: IBM]
{\small Correlation plot between the energy ratios $R_{0/2}$ and $R_{4/2}$: 
(left) experimental values \cite{Pietralla} and (right) theoretical values 
calculated in the consistent Q-formulation of the IBM.}
\label{norbert}
\end{figure}

Recently, an interesting empirical correlation was pointed out 
\cite{Pietralla} between $R_{4/2}$ of Eq.~(\ref{r42}) and
\ba
R_{0/2} \;=\; \frac{E(0^+_2)-E(0^+_1)}{E(2^+_1)-E(0^+_1)} ~. 
\ea
The left panel of Fig.~\ref{norbert} shows the experimental values for the 
nuclei with neutron number $84 \leq N \leq 96$ and the right panel the 
results of a IBM calculation in which the parameters of the CQF Hamiltonian 
of Eq.~(\ref{hcqf}) are taken randomly, but are restricted to the physically 
allowed region, {\it i.e.} $\epsilon > 0$, $\kappa > 0$ and 
$-\sqrt{7}/2 \leq \chi \leq \sqrt{7}/2$. For vibrational nuclei both energy 
ratios are equal to 2, since the $0^+_2$ and $4^+_1$ states belong to the 
two-phonon multiplet, whereas for rotational nuclei the $4^+_1$ state is a 
rotational excitation and the $0^+_2$ state a vibrational excitation 
(see Table~\ref{BE2}). Fig.~\ref{norbert} shows that this trend is satisfied
both by the data and by schematic IBM calculations in which only some 
minimal restrictions are imposed upon the Hamiltonian.

\subsection{Signatures of critical point nuclei} 

\begin{table}
\centering
\caption[]{\small Comparison of energy ratios for the critical points of 
the IBM Hamiltonians $H_{12}$ and $H_{13}$ with $N = 11$ and the critical 
point symmetries \protect\cite{QPT}.}
\label{critical}
\vspace{15pt}
\begin{tabular}{ccccc}
\hline
& & & & \\
& $\begin{array}{c} U(5) - SO(6) \\ \eta_c = 1/4 \end{array}$ & $E(5)$ 
& $\begin{array}{c} U(5) - SU(3) \\ \eta_c = 2/9 \end{array}$ & $X(5)$ \\
& & & & \\
\hline
& & & & \\
$R_{4/2}$ & 2.22 & 2.20 & 2.33 & 2.91 \\
$R_{0/2}$ & 2.61 & 3.03 & 2.41 & 5.67 \\
& & & & \\
\hline
\end{tabular}
\end{table}

Finally, I briefly address the question of experimental signatures of 
critical point nuclei \cite{RFC}. Phase transitions in nuclei can be tested 
experimentally by measuring observables that are particularly sensitive 
to them, such as separation energies $S_{2n}$, isomer shifts 
$\delta \langle r^2 \rangle$ and the transition rates 
$B(E2; 2^+_3 \rightarrow 0^+_2 )$ \cite{IZC}. Other observables that have 
been suggested as measures of critical point nuclei are the energy 
ratios $R_{4/2}$ and $R_{0/2}$. Table~\ref{critical} shows the values 
of these ratios for the IBM Hamiltonians $H_{12}$ and $H_{13}$ at the 
critical points $\eta_c = 1/4$ and $\eta_c = 2/9$, respectively. 
A comparison with the values for the critical point symmetries, $E(5)$ and 
$X(5)$ \cite{QPT}, shows that the energy ratios for the critical points 
in the IBM are rather different from the those for the critical point 
symmetries. This is not so surprising, since the potential-energy surfaces 
for IBM Hamiltonians at the critical points are not square-well potentials 
(see Fig.~\ref{pot1}) as in the critical point symmetries \cite{QPT}. 
Moreover, one has to take into account that there is an entire class of 
IBM Hamiltonians which all share the same potential-energy surface and 
hence have the same critical points, but differ in their kinetic energies. 
Therefore, there is not a single value of the energy ratios $R_{4/2}$ 
and $R_{0/2}$ that can be used to identify a nucleus as a "critical point" 
nucleus. Instead one should rather study the change in energy ratios, 
$B(E2)$ values and/or quadrupole moments in a chain of nuclei to determine 
whether these nuclei exhibit a shape-phase transition.

\section{Summary, conclusions and outlook}

In these lecture notes I have presented a review of shape-phase transitions 
in nuclei and random interactions. Maybe at first sight, these two topics 
have little to do with one another, but both were shown to provide 
evidence for the existence of robust shapes in collective nuclei. The 
transitions between different shapes occur very rapidly, typically with 
the addition or removal of one or two pairs of nucleons. 

An analysis in IBM-1 and IBM-2 showed that the phase transtions between 
spherical and axially deformed nuclei are of first order, whereas those 
between spherical and triaxially deformed and between axially deformed 
and triaxially deformed are of second order.

Two-neutron transfer reactions $(p,t)$ and $(t,p)$ for the Ge isotopes were 
shown to provide a sensitive probe of shape-phase coexistence of a normal 
and an intruder configuration. In particular, the ratio of cross sections 
for the excitation of the first excited $0^+$ state and the ground state is 
either very small or zero in the absence of configuration mixing, but may 
be enhanced for strongly mixed configurations. 

The importance of the concept of geometric shapes in nuclei was confirmed 
by a study of the IBM with random interactions. Despite the random nature 
of the interactions a highly surprising degree of regularity was observed, 
in particular the dominance of ground states with $L = 0$ and the occurrence 
of vibrational and rotational band structures. It was shown that the 
geometric shapes associated with IBM Hamiltonians play a crucial role 
in understanding the origin of these regular properties. These shapes are 
a reflection of an intrinsic geometry associated to the many-body dynamics 
of the model space which is sampled by the statistical nature of the random 
interactions, but which is quite independent of them. The observed spectral 
order is a robust property that arises from a much larger class of 
Hamiltonians than is usually thought.

\section*{Acknowledgments}

It is a pleasure to thank Pepe Arias, Jos\'e Barea, Mark Caprio, Octavio 
Casta\~nos, Rick Casten, Alejandro Frank, Franco Iachello, Elizabeth Padilla 
and Norbert Pietralla for interesting discussions. This work was supported 
in part by CONACyT.

\end{document}